\documentclass[journal]{IEEEtran}

\usepackage{cite}
\usepackage{array}
\usepackage{color}
\usepackage{float}
\usepackage[cmex10]{amsmath}
\usepackage{nomencl}						
\usepackage[normalem]{ulem}			
\usepackage{diagbox}						
\usepackage{slashbox}						
\usepackage{colortbl}						
\usepackage{multirow}						
\usepackage{tabularx}
\usepackage{placeins}
\usepackage{algorithm}
\usepackage[noend]{algpseudocode}
\usepackage{siunitx}
\usepackage{gensymb}
\usepackage{subcaption}

\ifCLASSINFOpdf
  \usepackage[pdftex]{graphicx}
	\graphicspath{{./figure/}}
  \DeclareGraphicsExtensions{.pdf,.jpeg,.png}
\else
  \usepackage[dvips]{graphicx}
  \graphicspath{{./figure/}}
  \DeclareGraphicsExtensions{.eps}
\fi

\begin{document}
\bstctlcite{IEEEexample:BSTcontrol}
\setcounter{page}{1}

\title{Phase-Compensated Frequency-Scan Measurement of Harmonic Transfer Function Impedance}
\author{Yiyang Liao, Yitong Li, \IEEEmembership{Member, IEEE}, Jinjun Liu, \IEEEmembership{Fellow, IEEE}}

\ifCLASSOPTIONpeerreview
	\maketitle 
\else
	\maketitle
\fi


\begin{abstract}
Impedance models obtained through frequency scanning provides terminal characterization for small-signal stability analysis without requiring detailed knowledge of internal converter parameters. When unbalanced steady-state components introduce cross-frequency coupling, conventional $2\times2$ impedance scan methods must be extended to identify harmonic transfer function (HTF) impedance. However, the measured phases of frequency-coupling terms depend on the starting time of the fast Fourier transform (FFT) window, potentially introducing phase errors when different time references are used. This paper presents a $dq$-domain measurement method of HTF impedance that explicitly accounts for this time dependence. Phase compensation aligns the measured Fourier coefficients to a common electrical phase reference, correcting the resulting phase discrepancies and enabling consistent impedance reconstruction. Comparisons with analytical models under balanced and unbalanced conditions verify the measurement method, recovering the conventional $2\times2$ impedance and the dominant coupling blocks of a $6\times6$ HTF representation, respectively.
\end{abstract}




\section{Introduction}
\label{sec:htf_introduction}

With the increasing deployment of power electronic converters in modern power systems, small-signal stability analysis has become increasingly important for ensuring stable system operation. Impedance-based analysis provides a practical approach to assessing small-signal stability by representing individual components through their terminal voltage--current relationships \cite{Sun2009review,Sun2011criterion}. Since terminal impedances can be obtained through either analytical modeling or measurement, impedance-based analysis is applicable to both systems with known internal dynamics and black-box system with unknown internal structures.

Under balanced sinusoidal steady-state conditions, the small-signal terminal dynamics of three-phase converters can commonly be represented by a $2\times2$ linear time-invariant (LTI) impedance matrix in a synchronous $dq$ frame \cite{Wen2016DQ}. Equivalent complex-vector representations have also been established \cite{Harnefors2007complex,Li2021circuit}. When unbalanced or harmonic steady-state components are retained, periodic coefficients may remain in the linearized model, leading to linear time-periodic (LTP) dynamics and coupling between different frequencies. Harmonic transfer functions (HTFs) describe these interactions through an infinite-dimensional transfer matrix whose elements relate the input and output harmonic channels \cite{Sandberg2005LTP}. HTF-based impedance and multifrequency admittance models have subsequently been developed to characterize the coupled dynamics of converters operating under unbalanced conditions \cite{Zhang2019HTF,Liao2020multifrequency,Guo2023HTF}. Li et al. then clarified how a low-dimensional representation can be recovered through HTF diagonalization under appropriate conditions \cite{Li2020interpreting}. 

Alongside analytical modeling, impedance measurement provides terminal characterization without requiring detailed knowledge of internal controller parameters. Conventional $dq$-domain frequency scans reconstruct the $2\times2$ impedance matrix under balanced conditions using two linearly independent perturbation tests. Practical implementations include line-to-line current injection \cite{Huang2009DQmeasurement} and perturbation injection using converters, induction machines, and chopper circuits \cite{Familiant2009measurement}. Beyond single-frequency scanning, time-domain multiple-input multiple-output (MIMO) parametric identification enables impedance estimation using simultaneous wideband excitation \cite{Gong2021MIMOmeasurement}. These methods provide useful foundations for terminal identification, while their extension to LTP systems requires resolving the additional coupled frequency channels.

Within the LTP framework, harmonic impedance matrices have been identified through small-signal current injections for stability assessment \cite{Salis2017HTFmeasurement,Salis2019HTFmeasurement}. However, the measured phase of a frequency-coupling term depends on the time origin of the fast Fourier transform (FFT), as discussed in stationary-frame and sequence-domain $2\times2$ impedance measurements \cite{Liao2020stationary,Shah2022sequence,Pedra2024review}. Specifically, changing the starting time of the FFT window changes the phases of the extracted voltage and current coefficients according to their respective frequencies. For a same-frequency impedance term, these phase changes cancel when the voltage--current ratio is calculated. For a frequency-coupling term, the input and output occur at different frequencies, leaving a residual phase shift proportional to both their frequency difference and the change in the FFT time origin. Consequently, even for the same operating condition and steady-state response, different FFT starting times can produce the same coupling magnitude but different coupling phases. In an HTF scan, coefficients collected from separate perturbation tests may therefore have inconsistent phase references, preventing their direct assembly into a consistently referenced impedance matrix. Although phase-reference compensation has been addressed for two-dimensional stationary-frame and sequence-domain measurements, its extension to $dq$-domain HTF scanning requires coordinated treatment across all retained frequency channels and perturbation tests. A systematic procedure is therefore needed to integrate perturbation injection, Fourier-coefficient phase compensation, and impedance-matrix reconstruction.

Therefore, this paper presents a frequency-scan procedure for measuring HTF impedances in the complex-vector $dq$ domain. Modulated three-phase voltage perturbations provide two independent excitation directions for each retained harmonic channel. The resulting voltage and current Fourier coefficients are aligned to a common electrical phase reference and assembled into matrices for impedance reconstruction. Comparisons with analytical models examine the conventional $2\times2$ representation under balanced conditions and the dominant frequency-coupling blocks of a $6\times6$ HTF representation under unbalanced conditions.

This paper is organized as follows. Section~\ref{sec:htf_measurement} introduces the $dq$-domain HTF impedance representation and the associated frequency-scan procedure. Section~\ref{sec:htf_verification} presents verification and discussion under balanced and unbalanced conditions. Section~\ref{sec:htf_conclusions} concludes the paper.


\section{$dq$-Domain HTF Impedance Measurement}
\label{sec:htf_measurement}

\subsection{HTF Impedance Representation}

For a three-phase system without a zero-sequence terminal path, the small-signal voltage and current variables are expressed in a common synchronous $dq$ frame. For $x\in\{u,i\}$, define the complex-vector $dq$ variables $x_{dq\pm}$ as \cite{Harnefors2007complex}
\begin{equation}
{x}_{dq\pm}(t)
=
\begin{bmatrix}
{x}_{dq+}(t)\\
{x}_{dq-}(t)
\end{bmatrix}
=
\underbrace{
\begin{bmatrix}1&j\\1&-j\end{bmatrix}
}_{T_j}
\begin{bmatrix}
{x}_d(t)\\
{x}_q(t)
\end{bmatrix}.
\label{eq:complex_transformation}
\end{equation}
These two components provide a complex-vector representation of the terminal variables in the same reference frame.

For the unbalanced operating conditions considered here, some coefficients of the linearized model are periodic at $2\omega_0$, where $\omega_0=2\pi f_0$. Multiplication by a periodic coefficient can generate responses at frequencies different from the input frequency. Specifically, the harmonic term $\boldsymbol{P}_{2m}e^{j2m\omega_0t}$ shifts an input component by $2m\omega_0$, coupling the input channel $s_{2n}$ to the output channel $s_{2(n+m)}$. In contrast, an LTI subsystem changes only the amplitude and phase of a sinusoidal steady-state component, without changing its frequency \cite{Sandberg2005LTP,Li2020interpreting}. The coupled components therefore belong to the harmonic frequency family
\begin{equation}
s_{2n}=s+j2n\omega_0,\qquad n\in\boldsymbol{Z}.
\label{eq:harmonic_channels}
\end{equation}
Here, $\boldsymbol{Z}$ denotes the set of integers.

Their infinite-dimensional harmonic vector is defined as
\begin{equation}
\tilde{\boldsymbol{x}}(s)
=\begin{bmatrix}
    \vdots \\[2pt] \tilde{x}_{dq\pm}(s_{-2}) \\[2pt] \tilde{x}_{dq\pm}(s) \\[2pt]  \tilde{x}_{dq\pm}(s_2) \\ \vdots
\end{bmatrix}, \\
\quad x\in\{u,i\},
\label{eq:stacked_harmonic_vector}
\end{equation}
with the channels ordered by increasing $n$ and two complex $dq$ components per channel.

With the terminal current defined as positive into the inverter, the HTF impedance relationship is
\begin{equation}
\tilde{\boldsymbol{u}}(s)
=
\boldsymbol{Z}_{\mathrm{HTF}}(s)
\tilde{\boldsymbol{i}}(s).
\label{eq:infinite_htf_relation}
\end{equation}
Let $\mathbf{Z}_{dq\pm,2h}(s)$ denote the $2\times2$ transfer block from a current perturbation at $s$ to a voltage response at $s+j2h\omega_0$. The HTF entries are then
\begin{equation}
\left[
\boldsymbol{Z}_{\mathrm{HTF}}(s)
\right]_{\ell n}
=
\boldsymbol{Z}_{dq\pm,2(\ell-n)}(s_{2n}),
\qquad \ell,n\in\boldsymbol{Z}.
\label{eq:htf_block_index}
\end{equation}
Thus, the block-row and block-column indices identify the output and input frequency channels, respectively. The diagonal blocks describe same-frequency responses, while the off-diagonal blocks describe cross-frequency responses.

The HTF impedance is theoretically infinite-dimensional because a perturbation can produce responses across an infinite set of coupled frequency channels. In practical analysis and measurement, a truncated HTF is therefore adopted to obtain a finite-dimensional approximation. The truncation order is selected to balance approximation accuracy, computational complexity, and measurement effort \cite{Sandberg2005LTP,Liao2022truncation}. Specifically, for a truncation order $N$, the frequency $n=-N,\ldots,N$ gives
\begin{equation}
\tilde{\boldsymbol{u}}_N(s)
\approx
\boldsymbol{Z}_N(s)\tilde{\boldsymbol{i}}_N(s),
\label{eq:truncated_terminal_relation}
\end{equation}
where $\tilde{\boldsymbol{x}}_N(s)$ is obtained from \eqref{eq:stacked_harmonic_vector} by retaining only the components at $s_{-2N},\ldots,s_0,\ldots,s_{2N}$, and $\boldsymbol{Z}_N(s)$ is a $(2N+1)\times(2N+1)$ matrix.

For example, $N=1$ retains $s_{-2}$, $s$, and $s_2$, yielding a $6\times6$ matrix. Writing $\boldsymbol{Z}_h$ for $\boldsymbol{Z}_{dq\pm,h}$, its block structure is
\begin{equation}
\boldsymbol{Z}_1(s)=
\begin{bmatrix}
\boldsymbol{Z}_0(s_{-2})
& \boldsymbol{Z}_{-2}(s)
& \boldsymbol{Z}_{-4}(s_2)\\
\boldsymbol{Z}_2(s_{-2})
& \boldsymbol{Z}_0(s)
& \boldsymbol{Z}_{-2}(s_2)\\
\boldsymbol{Z}_4(s_{-2})
& \boldsymbol{Z}_2(s)
& \boldsymbol{Z}_0(s_2)
\end{bmatrix}.
\label{eq:htf_n1}
\end{equation}
Its central input-frequency column consists of $\boldsymbol{Z}_{dq\pm,-2}(s)$, $\boldsymbol{Z}_{dq\pm,0}(s)$, and $\boldsymbol{Z}_{dq\pm,2}(s)$, which characterize the voltage responses at $s_{-2}$, $s$, and $s_2$ to current perturbations at $s$.

In the LTI limit, all cross-frequency blocks vanish and the HTF becomes block diagonal. Accordingly, $N=0$ recovers the conventional $2\times2$ terminal impedance, expressed as $\boldsymbol{Z}_{dq\pm,0}(s)$.

\subsection{Frequency-Scan Procedure} 

The truncated HTF impedance can be measured by injecting small perturbations and recording the resulting terminal voltage and current responses. Conventional $dq$-domain $2\times2$ impedance measurement uses two linearly independent perturbations to identify the terminal impedance matrix \cite{Huang2009DQmeasurement}. For an LTP system, this principle can be extended to the retained harmonic-frequency channels, where the measured voltage and current coefficients are assembled into harmonic matrices to reconstruct the HTF impedance \cite{Salis2019HTFmeasurement}. However, when frequency coupling is present, the phase of the measured coupling terms additionally depends on the starting time of the FFT window, which is firstly explicitly identified and addressed in sequence-impedance frequency-scan measurements \cite{Pedra2024review,Shah2022sequence}.

\begin{figure}[htb!]
\centering
\includegraphics[scale=1]{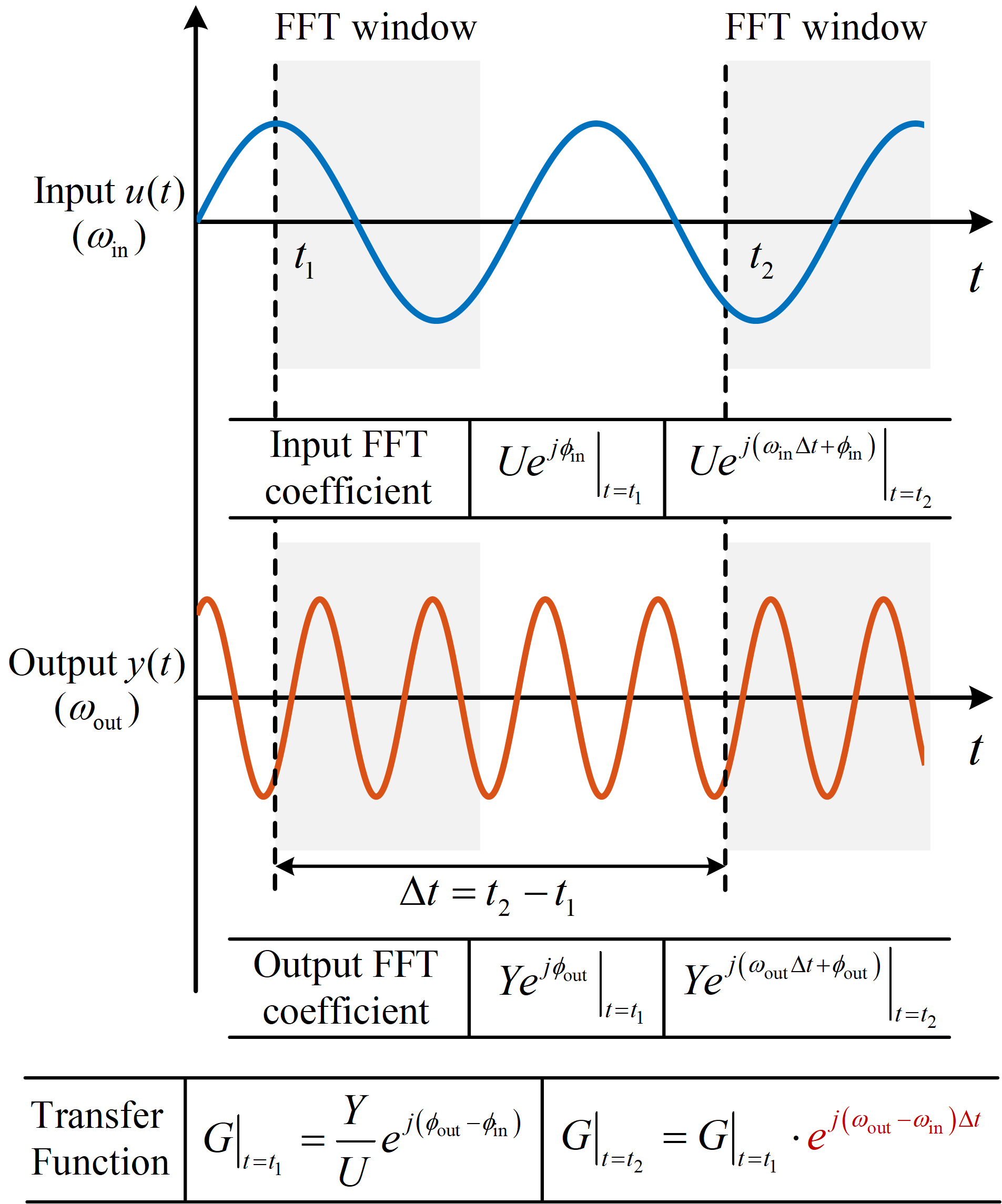}
\caption{FFT start time dependence of the frequency-coupling transfer function.}
\label{Fig:FFT_time_dependent}
\end{figure}

As illustrated in Fig. \ref{Fig:FFT_time_dependent}, shifting the FFT window from $t_1$ to $t_2$ introduces a phase shift in the measured frequency-coupling transfer function ($\omega_{\mathrm{in}}\neq\omega_{\mathrm{out}}$). In contrast,
same-frequency transfer terms ($\omega_{\mathrm{in}}=\omega_{\mathrm{out}}$) are invariant to the FFT-window starting point.

To eliminate this dependence, all Fourier coefficients are referred to a common phase reference \cite{Shah2022sequence}. Let $\alpha$ denote the phase of the phase-$a$ fundamental voltage at the beginning of the FFT window. A coefficient extracted at frequency $f$ is compensated as

\begin{equation}
X_{\mathrm{comp}}(f)
=
X(f)e^{-j(f/f_0)\alpha},
\label{eq:phase_compensation}
\end{equation}
where $f_0$ and $f$ denote the fundamental and extracted-component frequencies, respectively. This operation refers the measurement to an FFT window with zero phase of the phase-$a$ fundamental voltage (consistent with the theoretical reference $\theta_s(0)=0$).

\begin{figure}[htb!]
\centering
\includegraphics[scale=1]{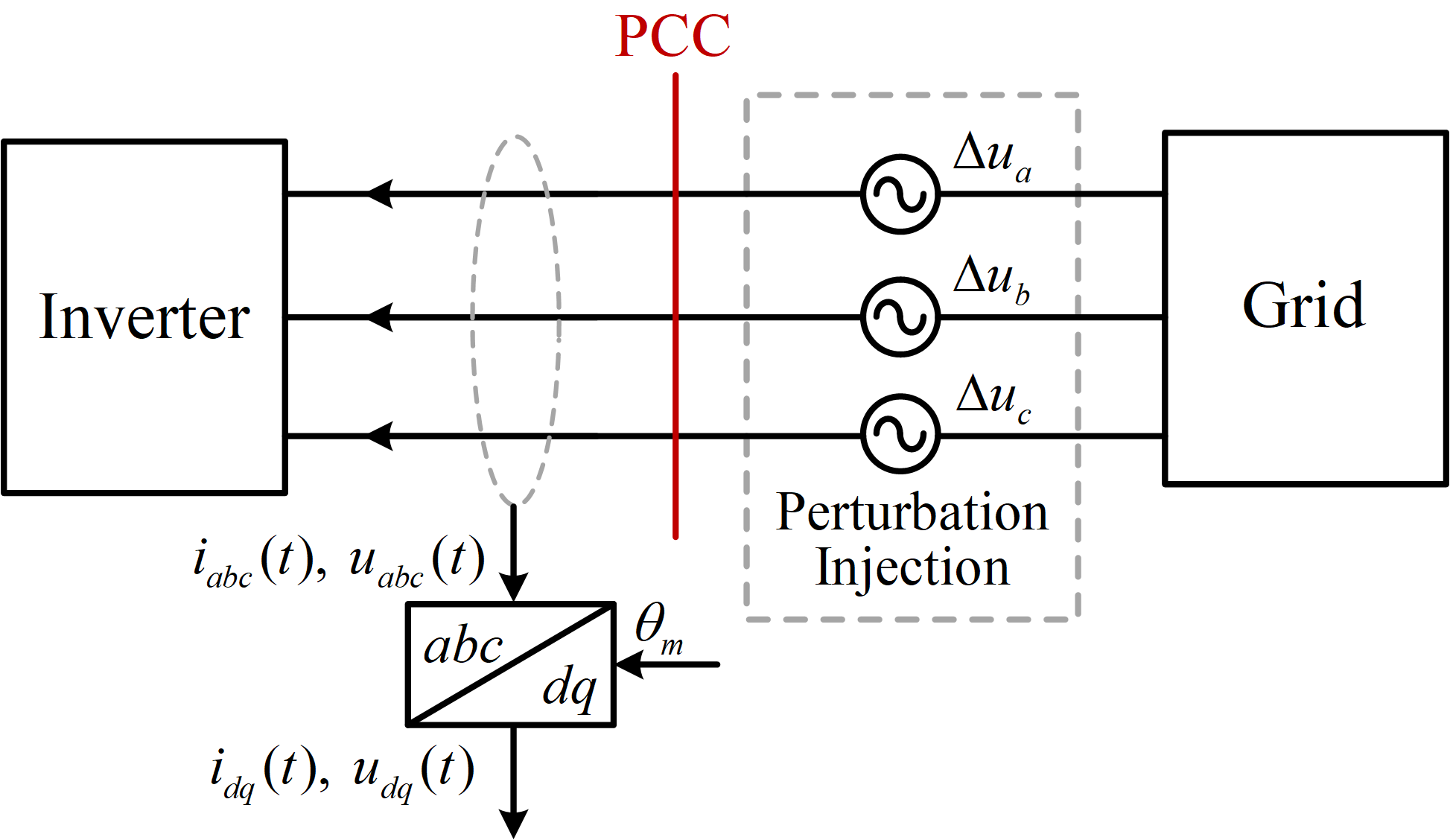}
\caption{Configuration of the proposed $dq$-domain HTF frequency-scan measurement.}
\label{Fig:Measurement_setup}
\end{figure}

The perturbation-injection configuration is shown in Fig. \ref{Fig:Measurement_setup}. Three series voltage sources are connected between the inverter and the grid, while the PCC voltages and currents are measured and transformed into the synchronous $dq$ frame using the angle $\theta_m$, which is usually obtained by an additional PLL.

Based on this configuration, the complete measurement procedure is summarized in Table \ref{Tab:HTF_frequency_scan_steps}. Briefly, modulated three-phase voltage perturbations are applied to excite the retained harmonic channels. The measured Fourier coefficients are then phase-compensated, transformed into complex $dq$ variables, and assembled to matrices form.

\begin{equation}
\begin{aligned}
    &\boldsymbol{U}_{dq\pm}(\omega_k)=
    \begin{bmatrix}
        U_{dq\pm}^{(-N)} & \cdots & U_{dq\pm}^{(0)} & \cdots & U_{dq\pm}^{(N)}
    \end{bmatrix} \\
    &\boldsymbol{I}_{dq\pm}(\omega_k)=
    \begin{bmatrix}
        I_{dq\pm}^{(-N)} & \cdots & I_{dq\pm}^{(0)} & \cdots & I_{dq\pm}^{(N)}
    \end{bmatrix}
\end{aligned}
\end{equation}
where $n=-N,\ldots,N$ indexes the injected frequency channel. Each $\boldsymbol{U}_{dq\pm}^{(n)}$ and $\boldsymbol{I}_{dq\pm}^{(n)}$ is a $2(2N+1)\times2$ block.

Therefore, the truncated HTF impedance at $f_k$ can be calculated.
\begin{equation}
    \boldsymbol{Z}_{\mathrm{HTF}}(\omega_k)=\boldsymbol{U}_{dq\pm}(\omega_k)[\boldsymbol{I}_{dq\pm}(\omega_k)]^{-1}
    \label{eq:Impedance_calculation}
\end{equation}

Repeating this procedure over the sweep-frequency sequence yields the complete frequency responses of the truncated HTF matrices.

\begin{table}[!htb]
\caption{STEPS OF THE PROPOSED IMPEDANCE FREQUENCY-SCAN METHOD}
\label{Tab:HTF_frequency_scan_steps}
\centering
\footnotesize 
\renewcommand{\baselinestretch}{1.1}\selectfont
\renewcommand{\arraystretch}{1.2} 
\setlength{\tabcolsep}{2.5pt} 

\renewcommand{\tabularxcolumn}[1]{m{#1}}

\begingroup
\begin{tabularx}{\columnwidth}{
    >{\centering\arraybackslash}m{0.11\columnwidth}|
    >{\raggedright\arraybackslash}X
}
\hline\hline
\textbf{Steps} & \multicolumn{1}{c}{\textbf{Actions}} \\
\hline

Step 1 &
Select the HTF truncation order $N$ and the sweep-frequency 
sequence $\boldsymbol{f}_{s}=\{f_1,\ldots,f_M\}$\\
\hline

Step 2 &
For the $k$th sweep frequency $f_k$, inject the following
modulated voltage perturbations over the retained harmonic channels:

{\centering
\par\vspace{5pt}
$\displaystyle
\left\{
\begin{aligned}
\Delta u_a(t)
&=
U_m\cos(2\pi f_{k,n}t)
\cos(2\pi f_0t-\phi_r)\\
\Delta u_b(t)
&=
U_m\cos(2\pi f_{k,n}t)
\cos\left(2\pi f_0t-\frac{2\pi}{3}-\phi_r\right)\\
\Delta u_c(t)
&=
U_m\cos(2\pi f_{k,n}t)
\cos\left(2\pi f_0t+\frac{2\pi}{3}-\phi_r\right)
\end{aligned}
\right.
$
\par\vspace{5pt}}

where $f_{k,n}=f_k+2nf_0, n=-N,\ldots,N$ indexes the retained harmonic channels
and $\phi_r=\{\phi_1,\phi_2\}$ represents the two injection angles.
\par\smallskip
\\
\hline

Step 3 &
Record the PCC voltage and current responses and extract Fourier coefficients at

{\centering
\par\smallskip
$\displaystyle
f_{k,\ell}=f_k+2\ell f_0,
\qquad \ell=-N,\ldots,N
$
\par\smallskip}
Apply the phase compensation in
\eqref{eq:phase_compensation}, followed by the complex
$dq$ transformation in \eqref{eq:complex_transformation}.
\\
\hline

Step 4 &
Assemble the voltage and current coefficient matrices and
calculate the HTF impedance at frequency $f_k$ according to \eqref{eq:Impedance_calculation}
\\
\hline

Step 5 &
Repeat Steps 2--4 for $k=1,\ldots,M$ and post-process
the measured HTF.
\\
\hline\hline

\end{tabularx}
\endgroup
\end{table}


\section{Verification and Discussion} 
\label{sec:htf_verification}

The proposed measurement procedure is evaluated using a grid-connected grid-forming inverter with droop control, and cascaded voltage and current control. Analytical impedance models obtained by linearizing the system under balanced and unbalanced operating conditions serve as references for the frequency-scan results. Under balanced conditions, the linearized terminal dynamics are LTI, and the HTF contains only block-diagonal terms. Accordingly, truncation with $N=0$ recovers the conventional $2\times2$ impedance without loss of terminal dynamics. Under unbalanced conditions, cross-frequency coupling arises, and $N=1$ is adopted to retain the dominant coupling channels, yielding a $6\times6$ HTF representation. Frequency scans are performed on the corresponding simulation model using the series voltage-injection. The measured PCC voltage and current responses are processed following Section \ref{sec:htf_measurement}, with the Fourier coefficients aligned to the same electrical phase reference as the analytical models.

\subsection{Balanced Conditions}

\begin{figure}[htb!]
\centering
\includegraphics[scale=1]{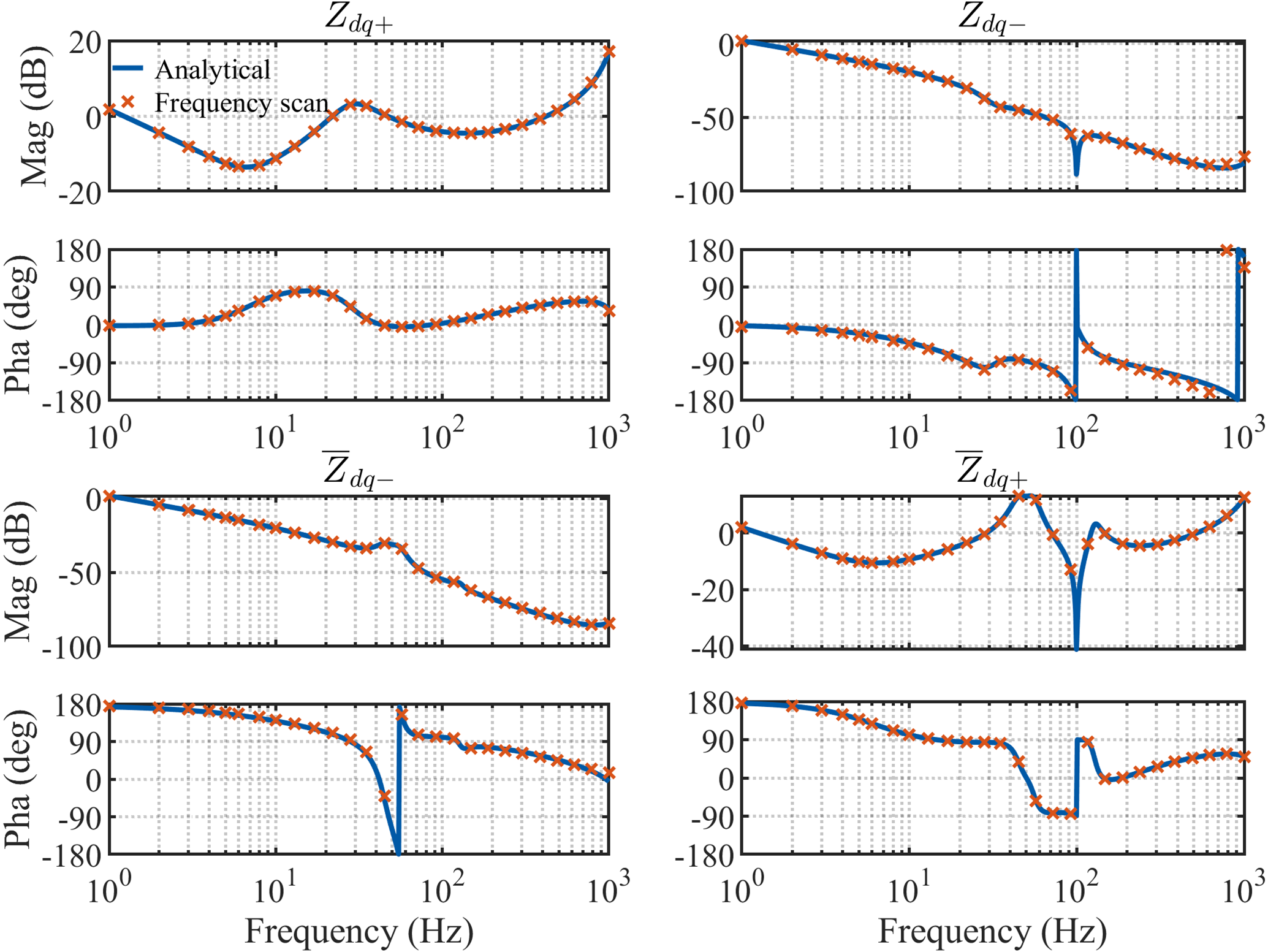}
\caption{Frequency-scan verification of the impedance model under balanced conditions.}
\label{Fig:Scan_balanced}
\end{figure}

The analytical impedance models are verified by comparison with the frequency-scan results. Under balanced conditions, the negative-sequence steady-state components vanish, and no cross-frequency coupling is present at the system terminals. The HTF therefore shows the block-diagonal characteristic and can be truncated with $N=0$, which retains only the central diagonal block and reduces the model to the conventional two-dimensional impedance. As shown in Fig.~\ref{Fig:Scan_balanced}, the analytical and frequency-scan results agree well over the frequency range of interest, confirming that the two-dimensional representation preserves the complete terminal dynamics under balanced conditions.

\subsection{Unbalanced Conditions}

Under unbalanced conditions, the negative-sequence steady-state trajectory introduces frequency coupling among harmonic channels separated by $2f_0$. Accordingly, the truncation order is selected as $N=1$, retaining the three frequency channels $f_k-2f_0$, $f_k$, and $f_k+2f_0$. Since each channel contains the forward and backward complex $dq$ variables, the resulting truncated HTF impedance is a $6\times6$ matrix. 

\begin{figure}[!htb]
    \centering
    
    \subfloat[$Z_{dq\pm,-2}(s)$.]{
        \includegraphics[scale=1]{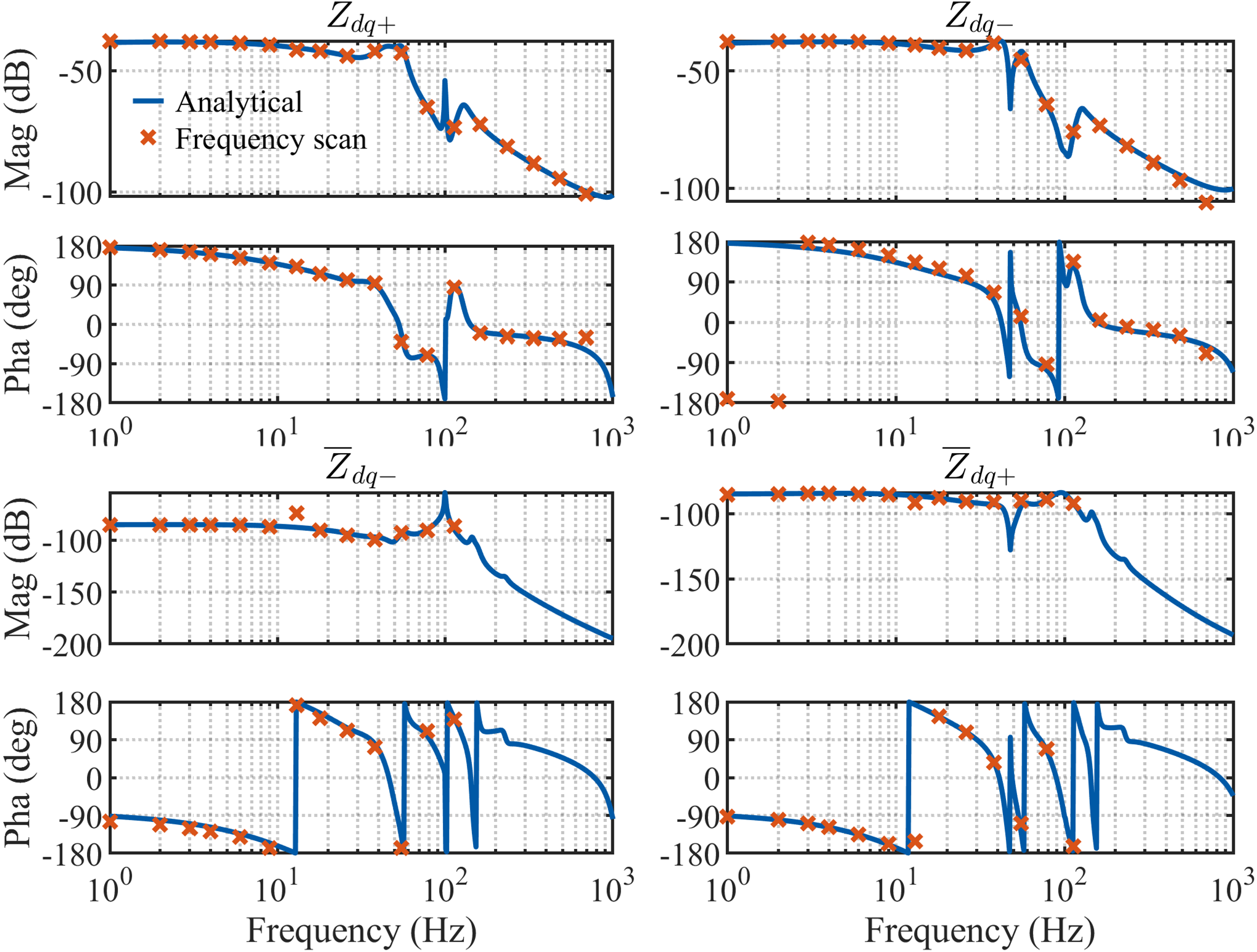}
    }\\
    
    \subfloat[$Z_{dq\pm,0}(s)$.]{
        \includegraphics[scale=1]{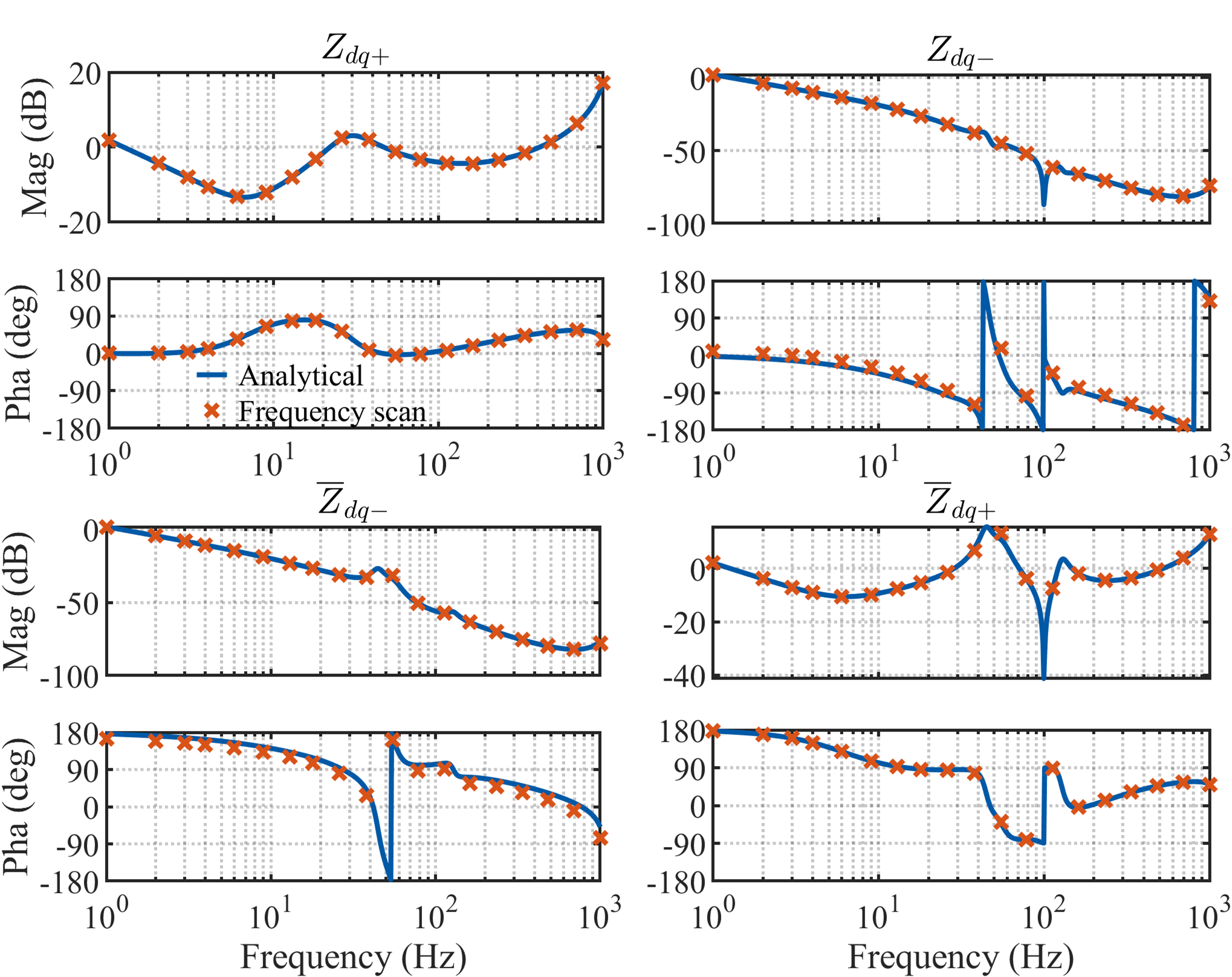}

    }\\
    
    \subfloat[$Z_{dq\pm,2}(s)$.]{
        \includegraphics[scale=1]{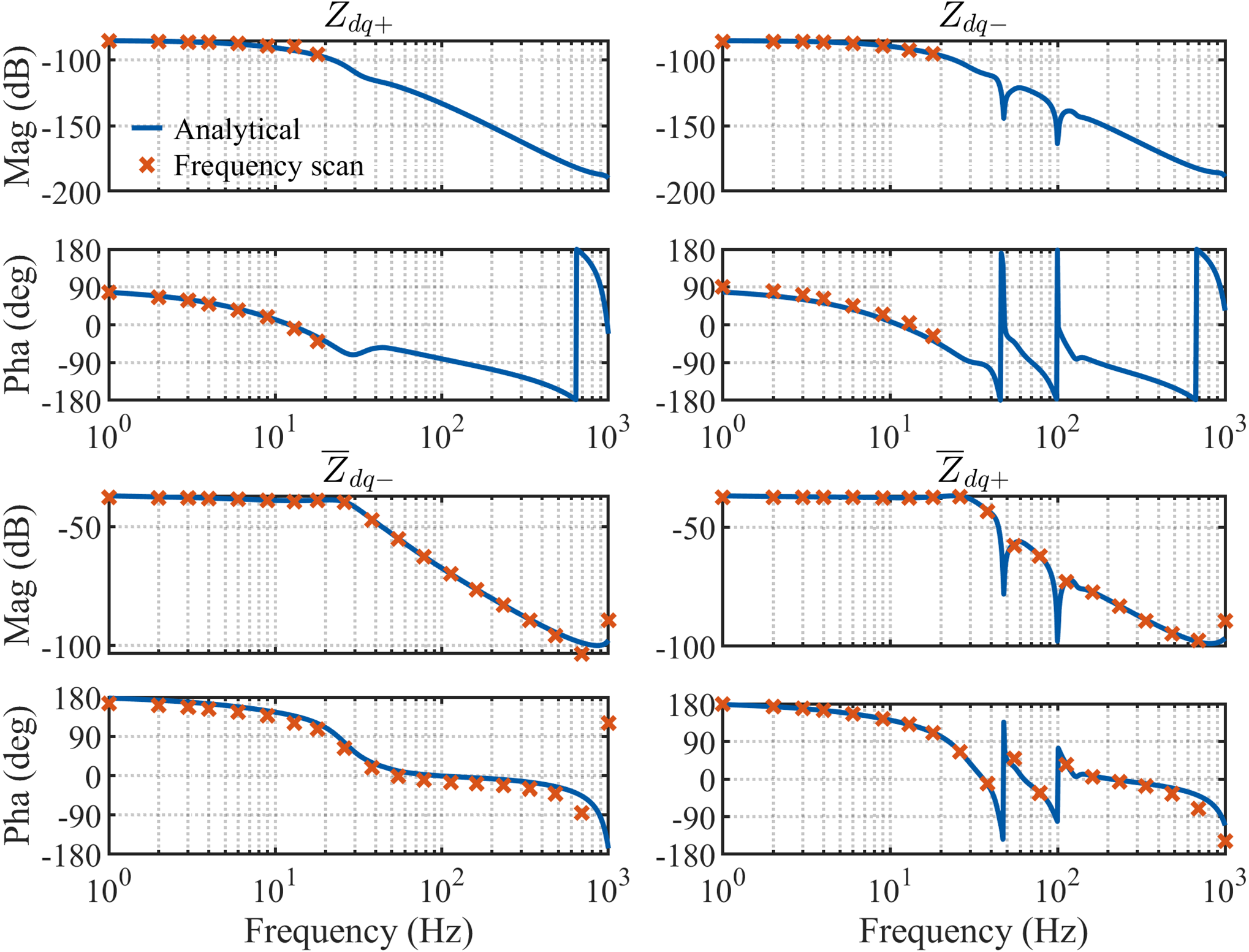}
    }
    
    \caption{Frequency-scan verification of the impedance model under unbalanced conditions.}
    \label{Fig:Scan_unbalanced}
\end{figure}

To avoid redundant presentation, Fig. \ref{Fig:Scan_unbalanced} shows only the central input-frequency column of \eqref{eq:htf_n1}, consisting of $\boldsymbol{Z}_{dq\pm,-2}(s)$, $\boldsymbol{Z}_{dq\pm,0}(s)$, and $\boldsymbol{Z}_{dq\pm,2}(s)$. The remaining blocks along the same block diagonals can be obtained through frequency shifting, whereas the higher-order corner blocks $\boldsymbol{Z}_{dq\pm,\pm4}$ have extremely small magnitudes and cannot be reliably identified through frequency scanning. Accordingly, frequency-scan results are omitted at frequencies where the corresponding analytical magnitude falls below $-100$ dB, as such weak responses cannot be reliably identified. The close agreement over the reliably identifiable frequency range validates the proposed HTF model.

\section{Conclusions} 
\label{sec:htf_conclusions}

This paper has presented a frequency-scan method for measuring truncated HTF impedances in the complex-vector $dq$ domain, explicitly accounting for the time-reference dependence of frequency-coupling terms. Since their input and output frequencies differ, changes in the FFT starting time introduce residual phase shifts that do not cancel in the measured transfer relationships. Referring the Fourier coefficients to a common electrical phase reference corrects these phase discrepancies and enables consistent HTF impedance reconstruction across frequency channels and perturbation tests. Comparisons with analytical models support the effectiveness of the method under balanced and unbalanced conditions. In the balanced case, the HTF is block diagonal, and $N=0$ recovers the conventional $2\times2$ impedance. In the unbalanced case, $N=1$ yields a $6\times6$ representation, with the dominant frequency-coupling blocks verified over the reliably identifiable frequency range. These results demonstrate the applicability of the method to both same-frequency and cross-frequency impedance measurement.



\ifCLASSOPTIONcaptionsoff
  \newpage
\fi

\bibliographystyle{IEEEtran}
\bibliography{Paper}

\end{document}